\documentclass[prb,reprint,superscriptaddress,showpacs]{revtex4-1}
\usepackage{graphicx}
\usepackage{amssymb,amsmath}
\usepackage{color}

\begin{document}

\title{
Lieb-Mattis ferrimagnetism in diluted magnetic semiconductors
}

\author{R.O.\ Kuzian} 
\affiliation{Institute for Problems of Materials Science
NASU, Krzhizhanovskogo 3, 03180 Kiev, Ukraine} 
\affiliation{Donostia International Physics Center (DIPC), ES-20018
Donostia-SanSebastian, Spain}

\author{J.\ Richter}
 \affiliation{Institut f\"ur Theoretische Physik,
 Otto-von-Guericke-Universit\"at Magdeburg,\\
PF 4120, D - 39016 Magdeburg, Germany}

\author{M.\ D.\ Kuz'min}
\affiliation{Aix-Marseille Universit\'e, IM2NP-CNRS UMR 7334, 
Campus St. J\'er\^ome, Case 142, 13397 Marseille, France}

\author{R. Hayn}
\affiliation{Aix-Marseille Universit\'e, IM2NP-CNRS UMR 7334, 
Campus St. J\'er\^ome, Case 142, 13397 Marseille, France}

\begin{abstract} 
We show the possibility of long-range ferrimagnetic ordering with 
a saturation magnetisation of $\sim 1\mu _B$ per spin for arbitrarily low 
concentration of magnetic impurities in semiconductors, provided that 
the impurities form a superstructure satisfying the conditions of the
Lieb-Mattis theorem. 
Explicit examples of such superstructures are given for the 
wurtzite lattice, and the temperature of ferrimagnetic transition is 
estimated from a high-temperature expansion. Exact diagonalization studies
show that small fragments of the structure exhibit enhanced magnetic 
response and isotropic superparamagnetism at low temperatures.
A quantum transition in a high magnetic field is considered and
similar superstructures in cubic semiconductors are discussed as well. 
\end{abstract}

\date{11.10.15} 

\pacs{75.10.-b,	
75.20.-g, 
   75.50.Gg, 
   75.50.Pp	
} 

\maketitle 

In order to launch the engineering of a new generation of electronic 
devices,
one needs new materials with special properties. For instance, 
spintronics
has a need for room-temperature ferromagnetic semiconductors
\cite{Zutic04}. Since the discovery of high-$T_C$ ferromagnetism in 
GaAs:Mn \cite{Matsukura98} and the prediction of room-temperature 
ferromagnetism in $p$-doped 
ZnO:Co,Mn systems \cite{Dietl00}, a lot of attempts have been made to obtain 
ferromagnetism in transition metal doped ZnO, GaN and in other oxides 
and nitrides. The $p$-type carriers doping is necessary for the $p$-$d$ Zener 
ferromagnetic long-range interaction \cite{Zener51}.  Up to now all  
attempts to obtain ZnO with $p$-type current carriers have failed. Nevertheless, 
several reports of ``ferromagnetic'' room temperature behavior 
have been published \cite{Dietl10,Janisch05,Ogale10}.  ``Perhaps the most 
surprising development of the past decade in the science of magnetic 
materials is the abundant observations of
spontaneous magnetization persisting to above room temperature
in semiconductors and oxides, in which no ferromagnetism
was expected at any temperature, particularly in the $p$-$d$ Zener
model'' \cite{Dietl10}.

In the absence of $p$-type current carriers, the interaction between 
magnetic impurities is governed by the superexchange mechanism. 
Superexchange is often regarded as an obstacle in the way towards magnetic 
semiconductors as it has antiferromagnetic (AFM) character and tends to 
anti-align the interacting spins, leading to a cancellation of the net
magnetization. In fact, the AFM interaction does \emph{not} preclude
spontaneous magnetization. In a seminal paper \cite{Lieb62},
E.\ Lieb and D.\ Mattis showed that the ground state of an AFM system 
depends on the topology of the interacting bonds and, under certain 
conditions, it is \emph{ferrimagnetic} rather than AFM.  
The Lieb-Mattis theorem applies if there is no magnetic frustration in the
spin system.

In this communication we study various structures formed by the interacting 
magnetic impurities in wurtzite semiconductors. We take antiferromagnetic 
nearest neighbor interaction into account and consider diluted 
lattices without frustration, in order to remain
within the Lieb-Mattis scheme. First we construct several finite 
clusters that show an enhanced magnetic response at low
temperatures.  Not alone do they possess a net magnetic moment, they all share a further 
interesting peculiarity: below a certain temperature their magnetic susceptibility
exceeds that of non-interacting spins. 
We call it isotropic superparamagnetic
response \cite{Bedanta09,Kuzian14b}. 
Next we construct extended lattices of these clusters, which undergo a 
ferrimagnetic ordering transition at a finite temperature. 
The average ground-state spin per magnetic ion of spin $S$ 
tends to a finite 
value (of about  $S/3$) despite the low concentration of magnetic ions.
The extension of our idea to other lattices and the influence of 
frustration will be briefly discussed at the end of the communication.

We take the interaction 
in the form
\begin{equation}
\hat{H}=\frac{1}{2}\sum_{\mathbf{R,r}}J_{\mathbf{r}}
\hat{\mathbf{S}}_{\mathbf{R}}\hat{\mathbf{S}}_{\mathbf{R+r}}, 
\label{Heis}
\end{equation}
i.e., we adopt the notation $J_{\mathbf{r}}$ for the interaction 
between one pair of spins \cite{remark1}.
We assume that only the 
nearest-neighbor (in the metal sublattice) interaction is nonzero. 
This assumption is relevant to magnetic semiconductors, where the 
nearest-neighbor exchange dominates\cite{Shapira02,Gratens04,DAmbrosio12}. 
Two kinds of nearest neighborships are present in wurtzites: those where both 
ions lie in the same plane and those where they lie in two adjacent planes.
The corresponding exchange integrals, $J_1$(in-plane) 
and $J_2$ (out-of-plane), are different \cite{Chanier06,Kuzian11,Savoyant14}.

\begin{figure}[htb]
\begin{minipage}[c]{.4\columnwidth}
 \includegraphics[width=0.8\textwidth]{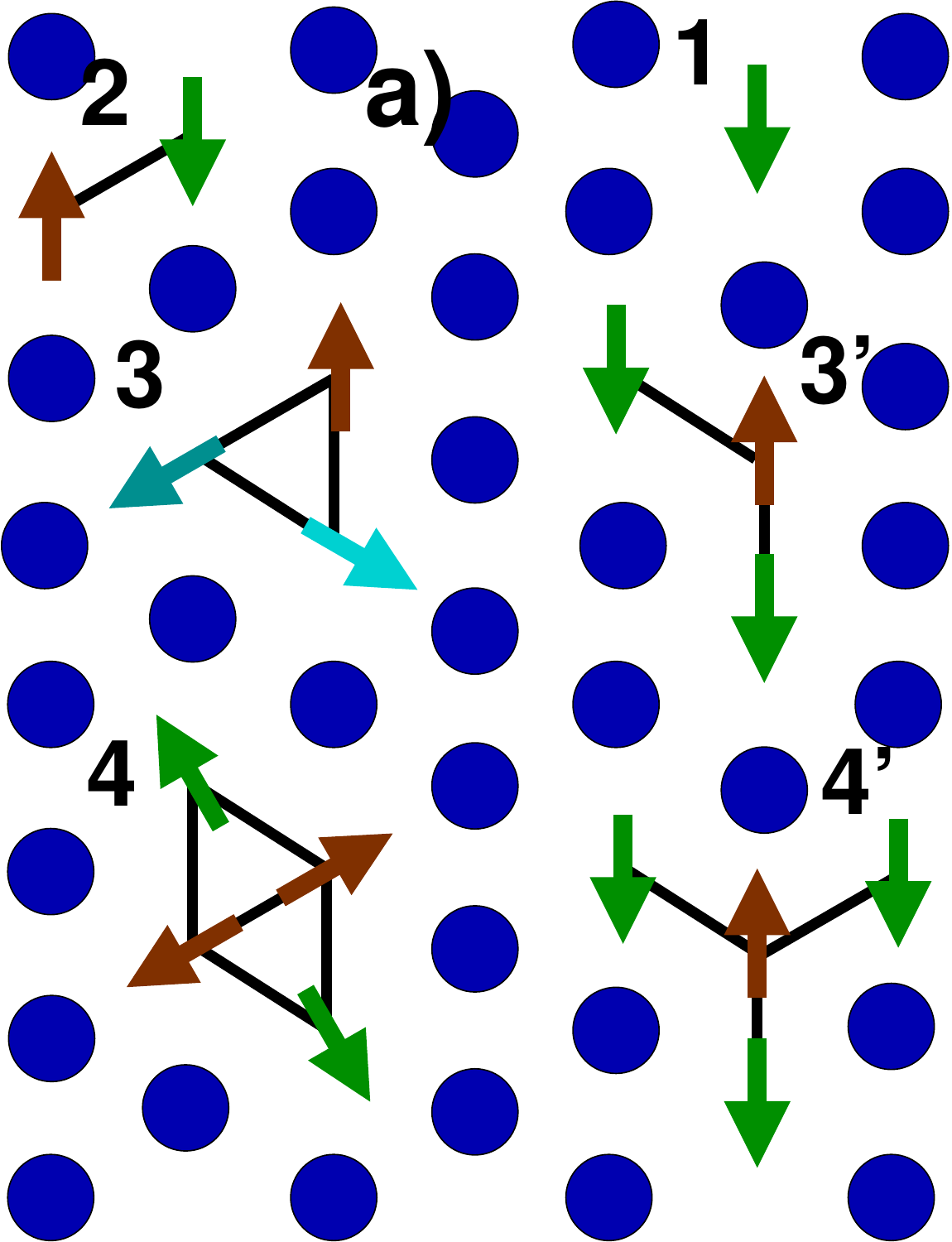}
\end{minipage}
\begin{minipage}[l]{.5\columnwidth}
 \includegraphics[width=0.7\textwidth]{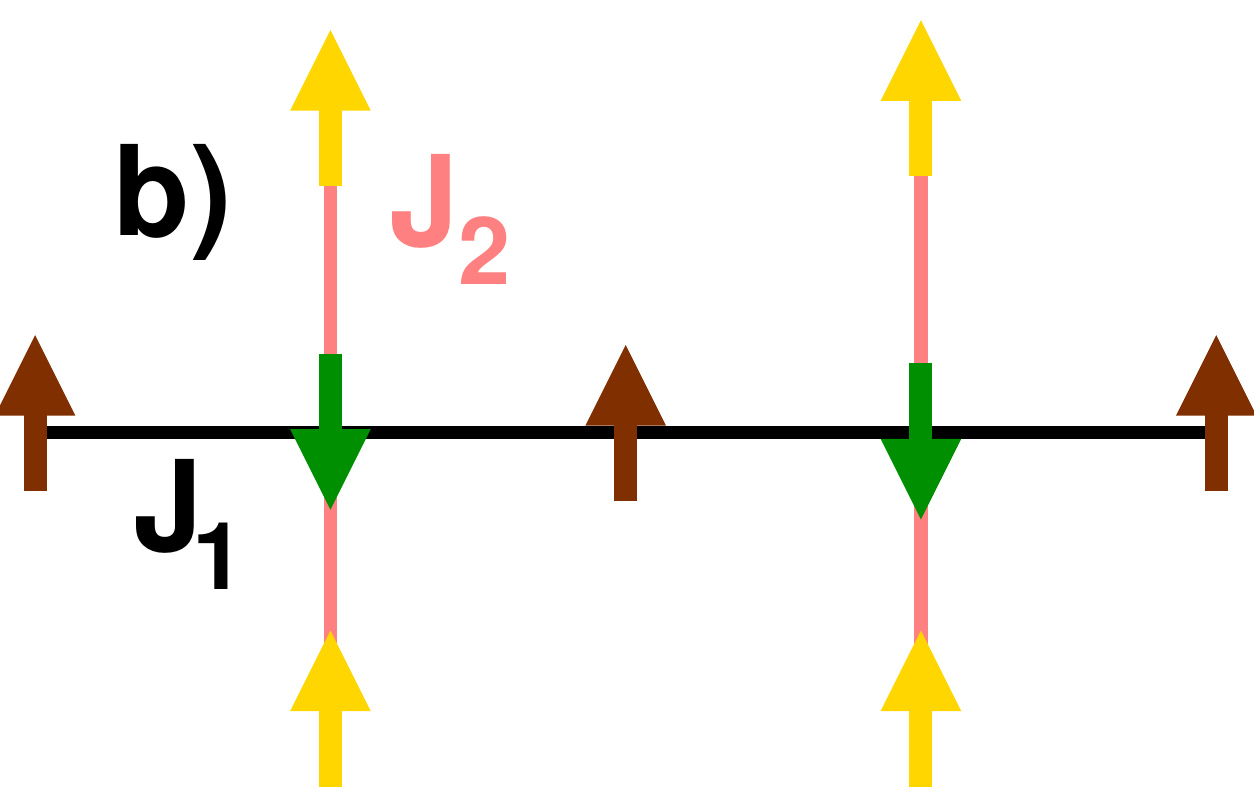}
 \includegraphics[width=\textwidth]{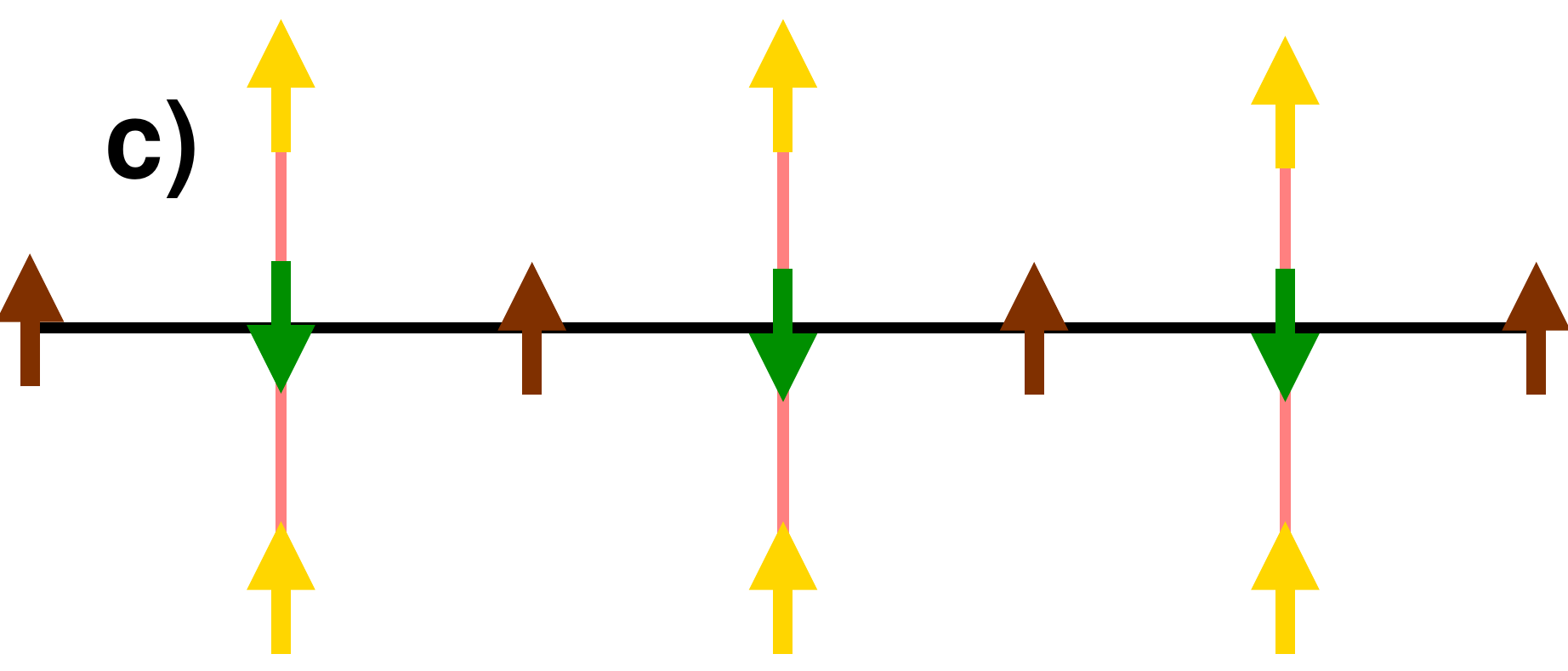}\\ 
\end{minipage}
\caption{(Color online) \textbf{a)} :
Complexes formed by transition metal impurities (arrows): isolated ions 
(1), dimers (2),  trimers (3,3$^\prime$), tetramers (4,4$^\prime$).  
Black solid line segments depict the nearest-neighbor interaction  $J_1$ bonds. 
One wurtzite $ab$ plane is shown, blue circles denote non-magnetic host 
metal ions, ligands are not shown.
\textbf{b), c)} : More complex  Lieb-Mattis systems with ferrimagnetic 
ground state: linear chains of impurities in 
the $ab$ plane "decorated" by spins in adjacent planes
(gold arrows);
pink line segments depict $J_2$ bonds.
}
\label{tetram}
\end{figure}

The magnetic response of a system is characterized by its magnetic susceptibility. 
Talking about a compound
 A$_{1-x}$M$_x$X (where X is a ligand of V or VI group, A is a metal of IIId or 
 IId group, and M is a transition metal), we shall attribute all the magnetic moment
 to transition metal ions (TMIs) only. 
 We now introduce the magnetic susceptibility per one spin, 
 \begin{equation}
 \chi   \equiv \frac{\mu_{M}}{H} , 
\end{equation}
where $\mu_M$ is the average magnetic moment of one TMI.
For non-interacting spins, the susceptibility obeys the Curie law
$\chi _C = [(g\mu _B)^2S(S+1)]/(3k_BT),$
where $S$ is the spin of the TMI and $g$ is its gyromagnetic ratio. Besides isolated 
spins, TMI impurities may form pairs, trimers, tetramers, and more complex 
structures (see Fig.\ \ref{tetram}).
The antiferromagneitc interaction depresses the magnetic response at high 
temperatures. For $T\gg J_{max}S(S+1)\equiv T_s$, the susceptibility of 
an interacting system obeys the Curie-Weiss law
$\chi _{CW} = [(g\mu _B)^2S(S+1)]/[3k_B(T-\theta)] < \chi _C $, with 
$-\theta =[S(S+1)]/(3k_B N)\sum_{\mathbf{R,r(R)}}J_{\mathbf{r(R)}}$.
Here $N$ is the number of spins and 
$J_{max}$ is the strongest exchange interaction in the system, 
$\mathbf{R}$ runs over all spins of the lattice, and $\mathbf{r}$
runs over all nearest neighbors of each spin.

\begin{figure}[htb]
\includegraphics[width=0.45\columnwidth]{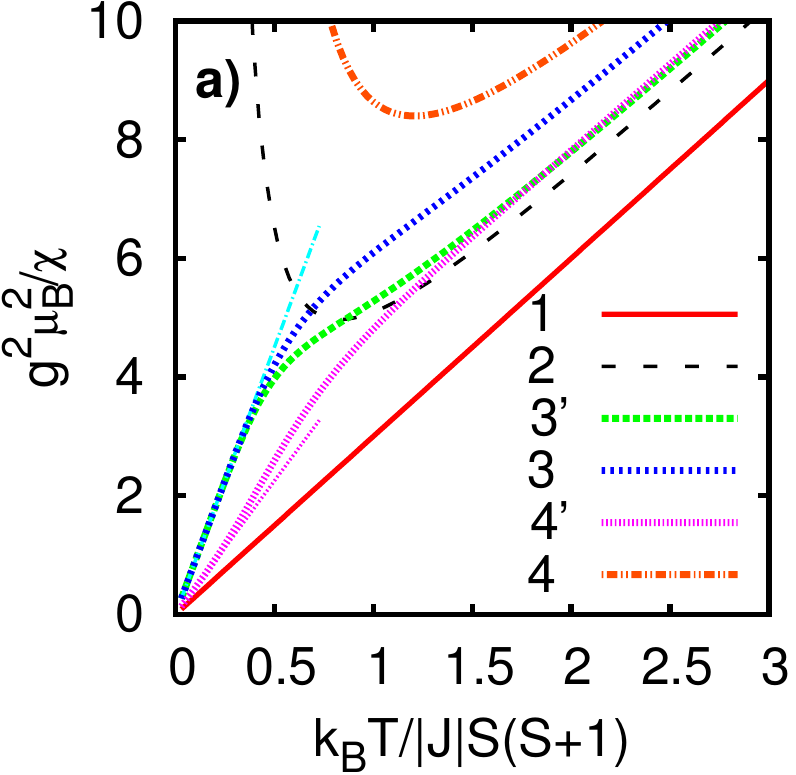}
\includegraphics[width=0.45\columnwidth]{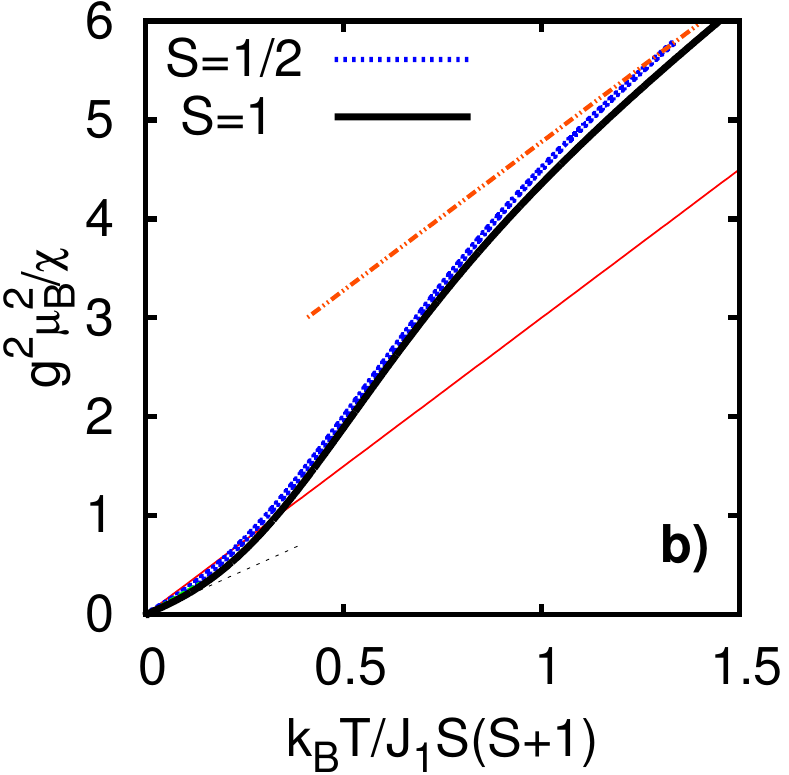}\\
\includegraphics[width=0.45\columnwidth]{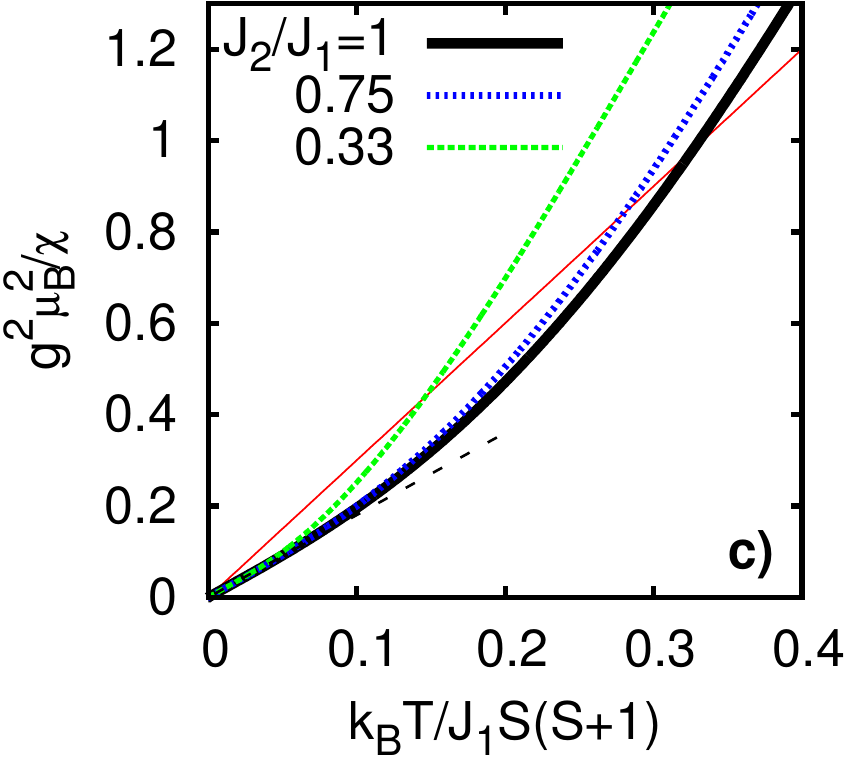}
\includegraphics[width=0.45\columnwidth]{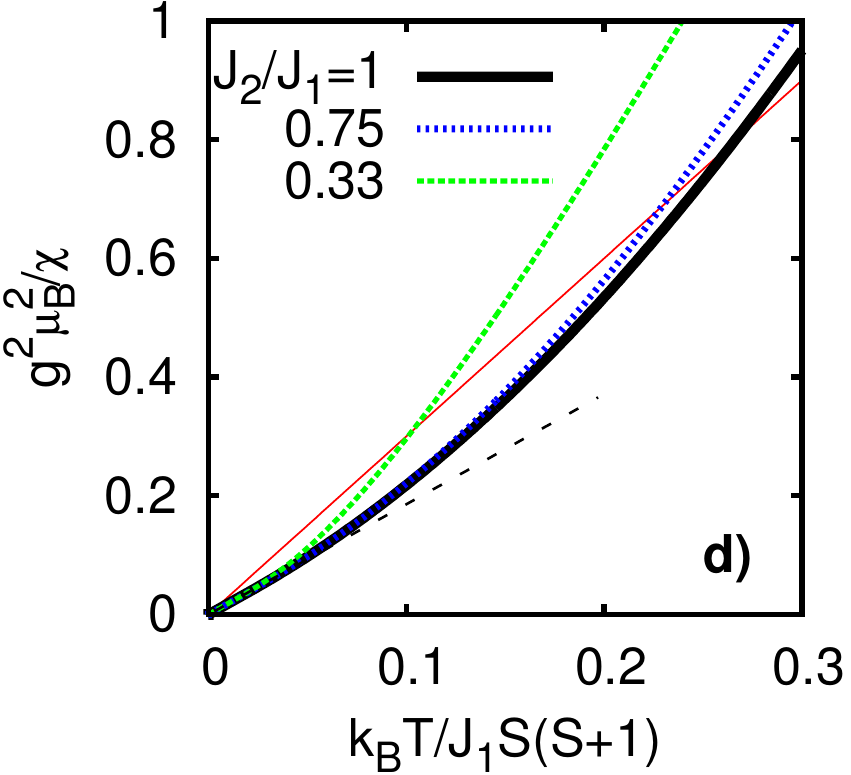}
\caption{(Color online) 
Inverse susceptibility (per spin) $\chi ^{-1}$ 
for the complexes shown in Fig.\ \ref{tetram}. 
Straight solid red line shows the Curie law  $\chi ^{-1}_C$;
straight dashed lines show the 
low-temperature asymptotics:  ``super'' -paramagnetic Curie laws 
$(g\mu _B)^2/\chi _g=3Nk_BT/[S_g(S_g+1)]$ 
for Lieb-Mattis systems.
\textbf{a)} : clusters shown in Fig.\ \ref{tetram}a with $S=1/2$;
\textbf{b)} : the complex shown in Fig.\ \ref{tetram}b with
two different  values of spin $S$; the straight dash-dotted red line is 
the high-$T$ Curie-Weiss asymptote. 
\textbf{c)} the same complex with $S=1$ and various values of $J_2/J_1$; 
\textbf{d)} the complex shown in Fig.\ \ref{tetram}c with $S=1/2$ and 
various values of $J_2/J_1$. 
}
\label{chif}
\end{figure}

At temperatures $T\lesssim T_s$, the 
response of the system depends on its geometry. Analytic expressions 
for the susceptibility can be obtained for small systems  \cite{Liu96}. 
Fig.\ \ref{chif}a shows the results for the simplest $S$=1/2 
case. 
We see that at $T\sim T_s$ the response of three spins arranged linearly 
\textbf{3}$^{\prime}$ is larger than that of 
a triangular arrangement of the same spins \textbf{3}. 
For 4-spin systems we see 
the striking difference between the response of a star arrangement 
\textbf{4}$^{\prime}$ 
and that of a rhombus \textbf{4}. 

Even more interesting is the response of the complexes shown in 
Fig.\ \ref{tetram}b,c. 
Each one of these systems can be decomposed into two sublattices A and B 
(denoted by arrows ``up'' and ``down''), the interaction being nonzero 
only between sites that belong to different sublattices. 
Such a system satisfies the 
requirements of the Lieb-Mattis theorem \cite{Lieb62}, and possesses 
a ferrimagnetic ground state with total spin $S_g=S|N_A-N_B|$. In this case, 
the term ``ferrimagnetic'' refers to correlations of the spins in the 
ground state, in the absence of a long-range magnetic order \cite{remark2}. 
We have performed full exact diagonalization studies (ED) of 
thermodynamic properties of clusters shown in 
Fig.~\ref{tetram}b,c using J. Schulenburg's {\it
spinpack} program \cite{spinpack1,spinpack2}. 
The susceptibility $\chi (T)$ is calculated as
the ratio of the induced magnetization $M$ to the 
"vanishing" magnetic field $H=10^{-5}J_1/g\mu _B$.
One observes in
Figure \ref{chif}b,c,d that the response of the systems 
shown in Fig.\ \ref{tetram}b,c \emph{exceeds} the response of 
non-interacting spins at low temperature. Thus, an antiferromagnetic 
interaction may result in an \emph{enhancement} of magnetic response if 
the geometry of spin arrangement favors the formation of a ferrimagnetic 
ground state. Then for temperatures $T\ll T_s$ the susceptibility per 
spin shows superparamagnetic response
$\chi _g = [(g\mu _B)^2S_g(S_g+1)]/[3k_BT(N_A+N_B)]$. Evidently, 
the enhancement of the low-temperature response 
takes place, if
\begin{equation}
K\equiv  \frac{\chi _g}{\chi _C} = 
\frac{|N_A-N_B|(|N_A-N_B|S+1)}{(N_A+N_B)(S+1)} >1. \label{K}
\end{equation}
Not every system satisfying the requirements of the Lieb-Mattis theorem
and having a ferrimagnetic ground state has an enhanced susceptibility.
Thus, the clusters $3^\prime$ $(N_A=1,N_B=2)$ and $4^\prime$ $(N_A=1,N_B=3)$
both have $K<1$, i.e. their response is weaker than that of the same number
of non-interacting spins.

The "S"-shape form of  the $T$-dependence of the inverse susceptibility 
(Figure \ref{chif}b) was previously reported for small fragments of ferrimagnetic 
superstructure in double perovskites \cite{Kuzian14b,Laguta14}. It  
interpolates between the Curie-Weiss law $\chi _{CW}$ at $T\gg T_s$, and 
the "super"-spin Curie law $\chi _g=K\chi _C$ at  $T\ll T_s$. 

If impurity spins arrange themselves in a periodic superstructure having 
two (or more) non-equivalent spin positions, 
a ferrimagnetic ground state is possible for this superstructure. 
Let us denote the number of spins in the superstructure unit cell 
$n_A+n_B$, where A and B refer to the non-equivalent positions. 
If the spins of the sublattice A interact (antiferromagnetically) 
only with the spins of the sublattice B (absence of frustration), 
and $n_A\neq n_B$ , the ground-state 
spin of the unit cell is $S_c=S|n_A-n_B|$ \cite{Lieb62}. 
\begin{figure}[htb]
\includegraphics[width=0.7\columnwidth]{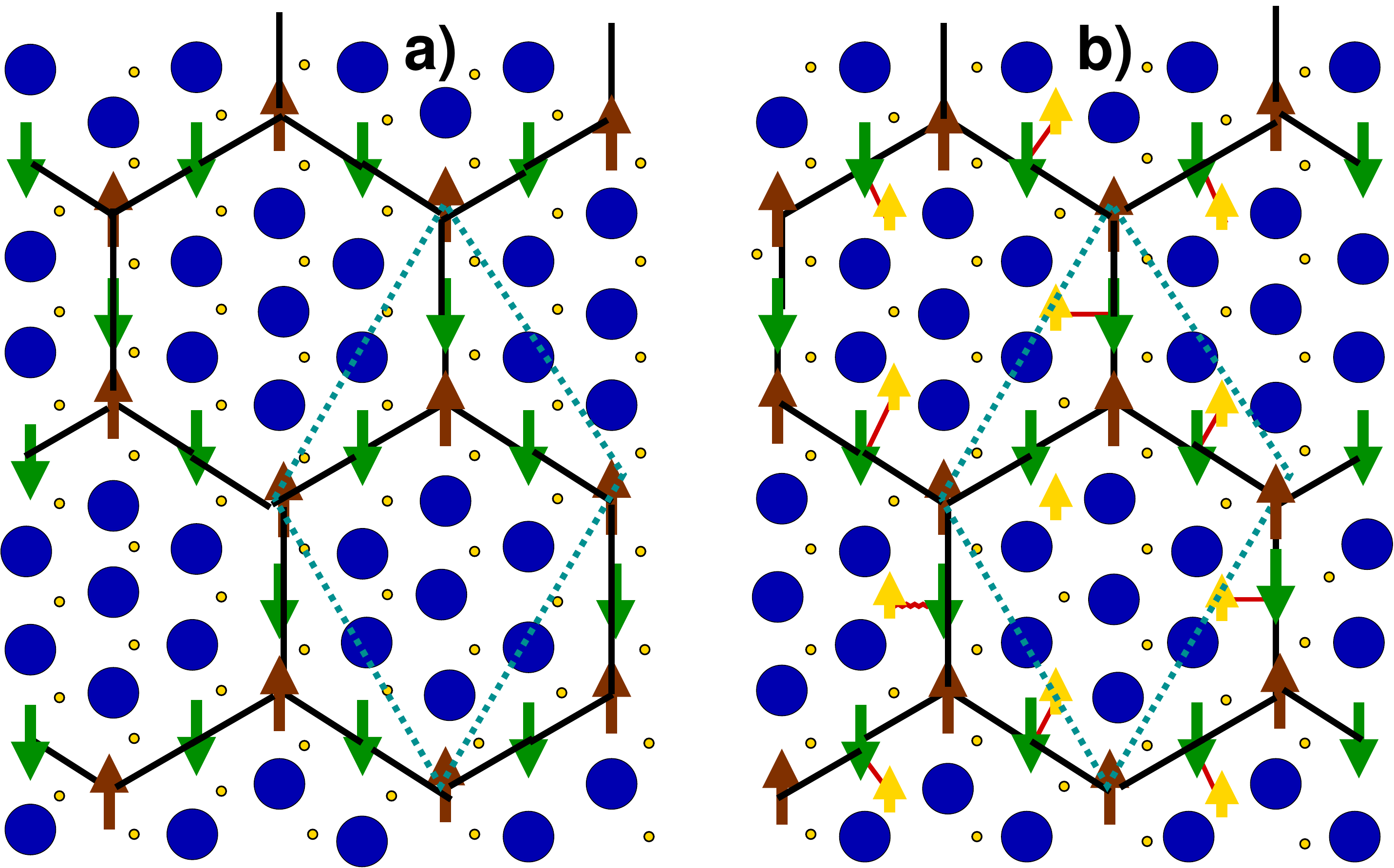}\\
\includegraphics[width=0.35\columnwidth]{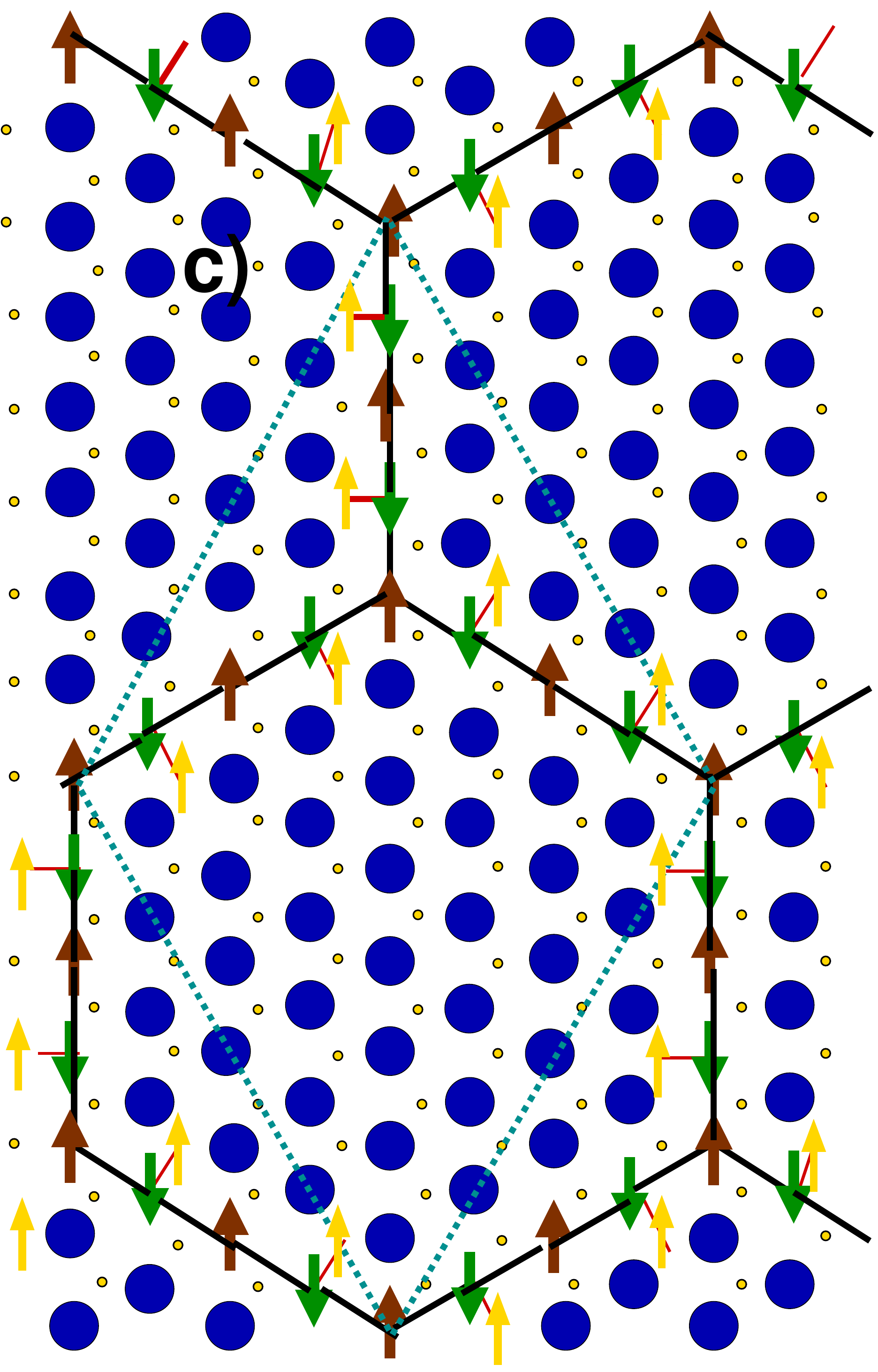}
\includegraphics[width=0.5\columnwidth]{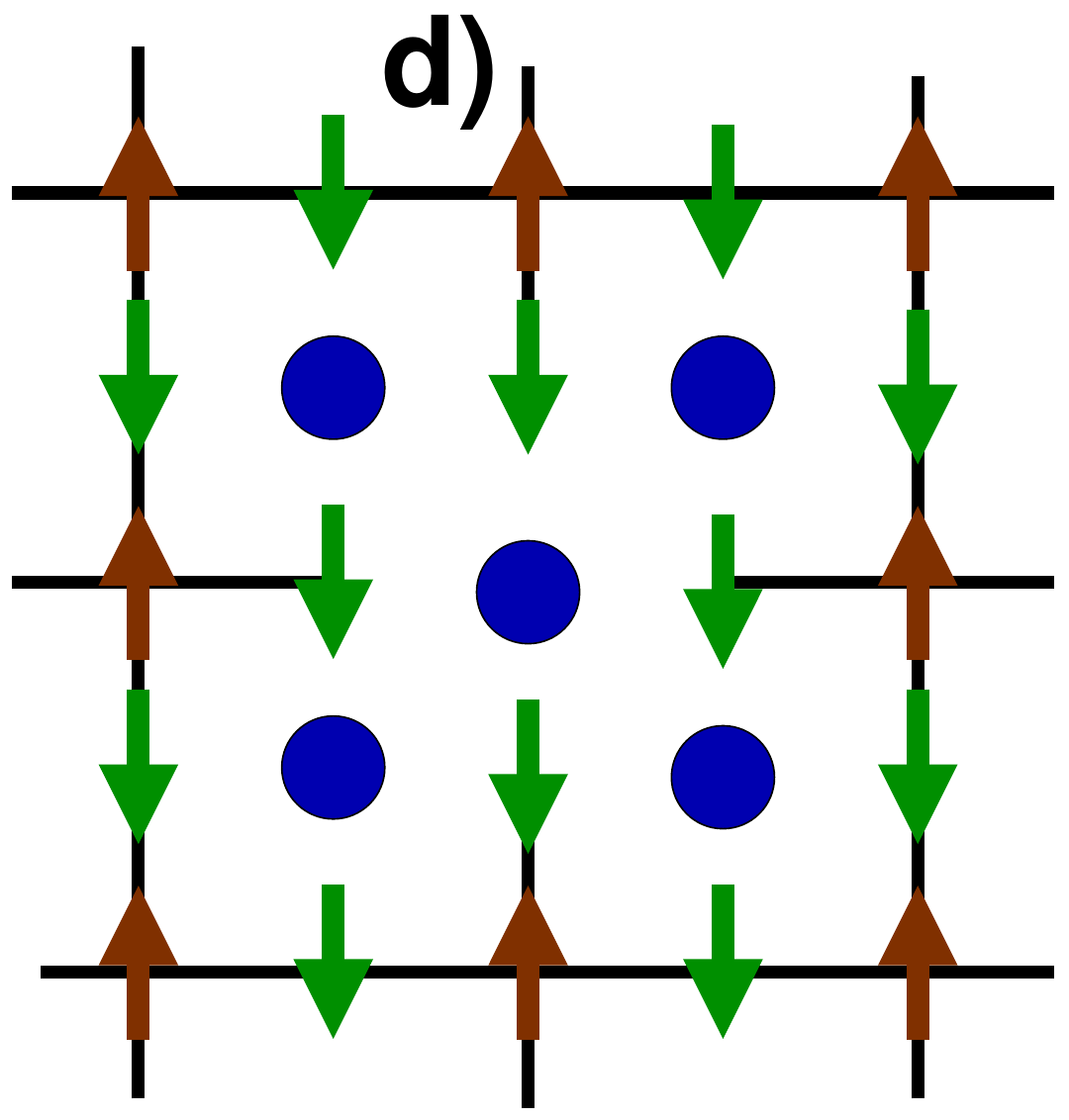}
\caption{(Color online) Examples of ferrimagnetic superstructures 
\textbf{ a), b) }: flat and three-dimensional two-leg honeycombs, $L=1$;
\textbf{ c) }: four-leg honeycomb, $L=2$;
\textbf{ d) }: a unit cell of a square network, it may be also 
regarded as a face of cubic unit cell.
The notations 
is the same as in Fig.\ \ref{tetram}.
The cyan rhombi show the unit cells.
}
\label{hon2d}
\end{figure}
For a fragment of such a ferrimagnetic superstructure containing $N_c$ cells, 
the ground-state spin is $S_g= N_cS_c=N_c|n_A-n_B|S$, and the enhancement 
ratio equals
$K = |n_A-n_B|(N_c|n_A-n_B|S+1)/[(n_A+n_B)(S+1)]$.
It is clear that for a sufficiently large number of cells $N_c$ the 
ratio $K$ will be not only greater than 1, but can reach very large 
values. Fig.\ \ref{hon2d}a shows a honeycomb superstructure 
that may be 
formed by TMIs in the $ab$ plane of the wurtzite structure. The hexagon 
edge length is
$a_h=2a$, $a$ being the lattice parameter of the wurtzite. 
It is easy to imagine superstructures with $a_h=2La$, $L=1,2...$, all of
them being ferrimagnetic.

Flat superstructures like those shown in Fig.\ \ref{hon2d}a can be linked together by 
some bridging spins to form a three-dimensional ferrimagnetic superstructure, 
which will undergo a ferrimagnetic phase transition, provided that the number 
of cells is macroscopically large. 
\begin{figure}[htb]
\includegraphics[width=0.45\columnwidth]{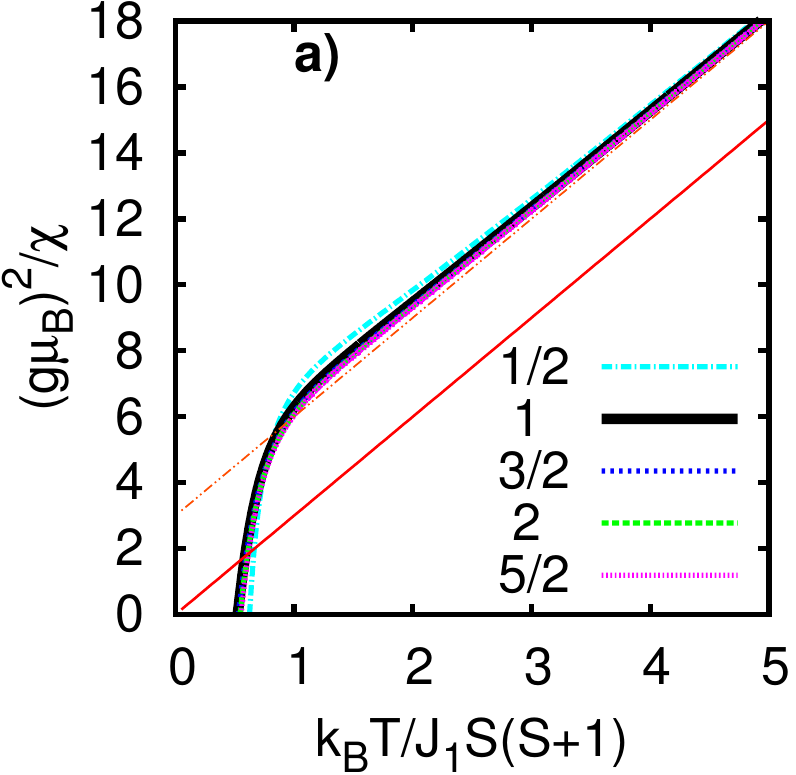}
\includegraphics[width=0.45\columnwidth]{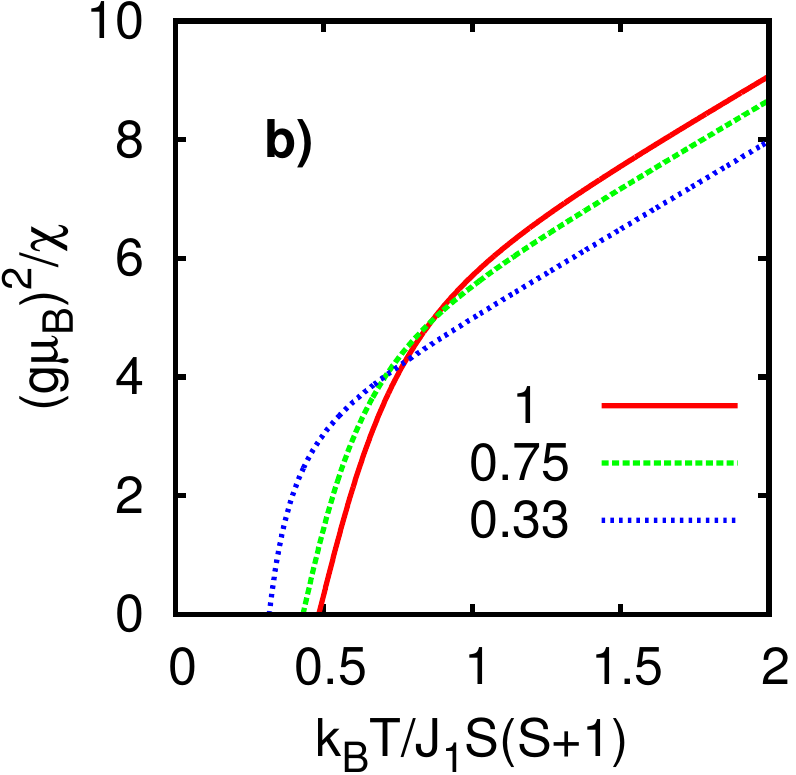}
\includegraphics[width=0.45\columnwidth]{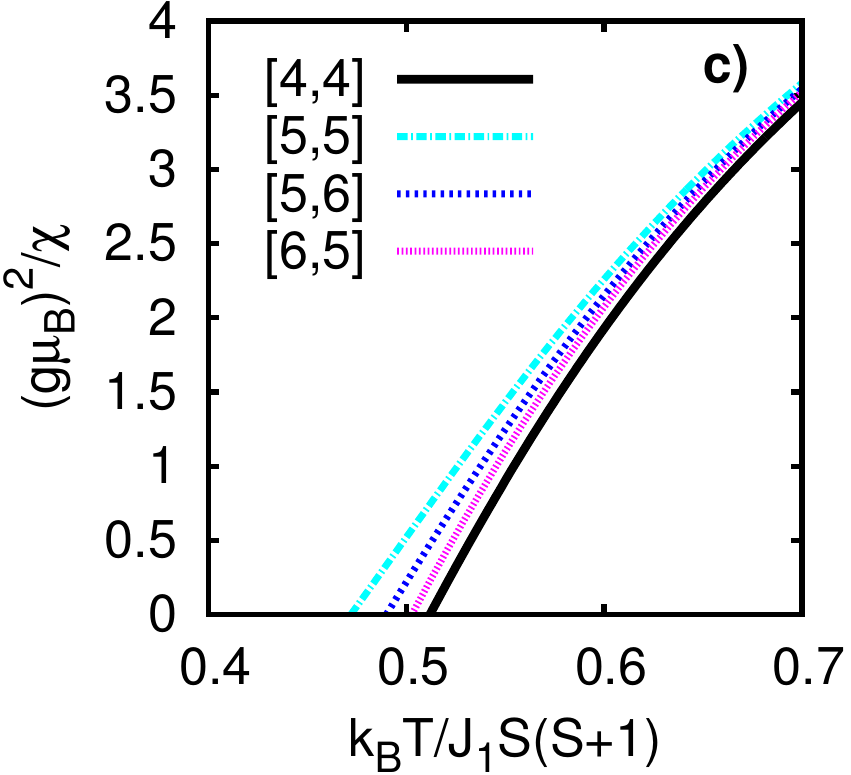}
\includegraphics[width=0.45\columnwidth]{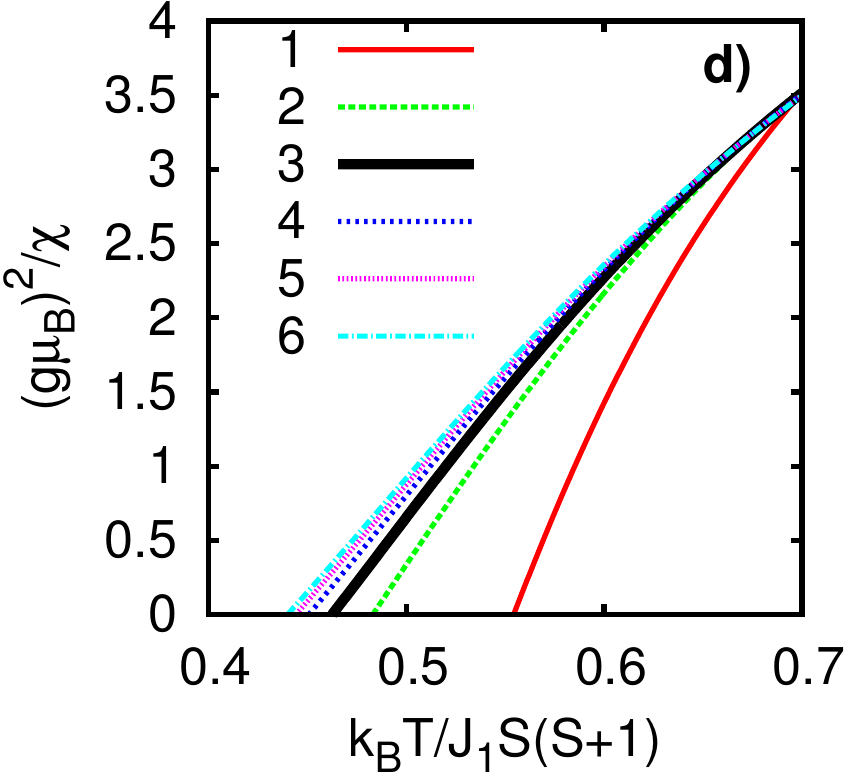}
\caption{(Color online) Temperature dependence of inverse susceptibility 
given by [5,5] Pad\'e approximants for tenth-order  high-temperature 
expansion (HTE) for ferrimagnetic superstructures: 
\textbf{ a) } two-leg honeycomb ($L=1$), 
various spin values are shown, solid (dash-dotted) straight red line
shows Curie (Curie-Weiss) law; 
\textbf{ b) } four-leg honeycomb ($L=2$), $S=5/2$ various 
$J_2/J_1$ values are shown; \textbf{ c) }  
four-leg system, $L=2$, $S=2$
various Pad\'e approximants for eighth-order 
([4,4])\cite{Schmidt11,hte8},
tenth-order([4,6], [5,5], [6,4]) \cite{Lohmann14}, 
and eleven-order ([5,6], [6,5]) \cite{Lohmann} HTE;
\textbf{ d) } the vicinity of $T_C$ for  various
honeycomb superstructures with size parameter $L=1,2\dots 6$, $S=5/2$, $J_2=J_1$.
}
\label{hte}
\end{figure}
Figure \ref{hon2d}b,c shows examples of the structures. It is clear that 
this motif may be repeated in an infinite number of variations. Like the 
host wurtzite lattice, the unit cell of the superstructure contains 
metal ions in two planes. The magnetic ions in one plane 
(green ``down'' and brown ``up'' arrows) form a honeycomb lattice with 
the hexagon edge $2La$. In the second plane, 
the magnetic ions (gold ``up'' arrows) occupy the positions nearest to 
the green ``down'' arrows. The interaction between the ions in the first plane 
is $J_1$, whereas the interaction between the ions in two adjacent planes is 
$J_2$. We note that the complexes shown in Fig.\ \ref{tetram}b,c are 
building blocks of the honeycombs. It will be demonstrated below that many other 
Lieb-Mattis networks can be built of such blocks.
The number of magnetic ions in the unit cell is $n_A+n_B=9L-1$ , 
the ground state spin of the cell being $S_c=S|n_A-n_B|=S(3L-1)$. Now the total number 
of ions in the cell is $n_c=24L^2$. Thus, the concentration of magnetic ions equals 
$x=(9L-1)/(24L^2)$, and can be made very small for sufficiently large $L$. 
At the same time, the average ground-state spin per magnetic ion, 
$\langle S_{\mathbf{R}}\rangle =S_c/(n_A+n_B)=S(3L-1)/(9L-1)$, tends to 
a finite value, $S/3$, as $L \to \infty$.

The inverse magnetic susceptibility $\chi ^{-1}$ 
of such superstructures is presented in Fig.\ \ref{hte} as a function of normalized
temperature $T/T_s$. 
It was calculated 
using a program \cite{Lohmann14} based on 
the tenth-order high-temperature expansion (HTE) \cite{hte10}. 
The program computes the exact coefficients of the HTE as well as its Pad\'{e}
approximants (ratios of two polynomials), 
$\chi (T) \approx [m,n]=P_m(T)/P_n(T)$. 
The Pad\'{e} approximants allow
to extend the region of validity of the HTE down to
$T\sim 0.5T_s$ \cite{Lohmann14} (Fig.\ \ref{hte}c). 
This extension sometimes fails if an  approximant has a pole in the 
temperature region of interest. Our experience shows that
the [5,5]  approximant works well in almost all cases.
Sometimes difficulties 
arise for $S=1/2$, and for small $J_2/J_1$ ratios, 
i.e., for the extreme 
quantum case. Nevertheless, due to the weak dependence of the shape of the 
curve $\chi^{-1}(T/T_s)$ on the spin value  $S$ (Fig.~\ref{hte}a), 
it can still be analyzed. 
At $T\gtrsim 3T_s$, the inverse susceptibility follows the Curie-Weiss asymptotic
 law with 
$\theta=-[S(S+1)/3k_B]12L(J_1+J_2)/(9L-1)$. For $T\lesssim T_s$ it 
sharply deviates from the asymptotic behavior and changes sign at $T=T_C$. 
This is the temperature of ferrimagnetic ordering --- the Curie temperature.

The  precision of the determination of critical temperatures 
from the zero of $\chi ^{-1}$ (Fig.\ \ref{hte}c) was estimated to be 
about 10\% \cite{Lohmann14}. 
Figure \ref{hte}b shows that $T_C$ decreases as 
the ratio of out-of-plane to in-plane 
couplings, $J_2/J_1$, is reduced. At $J_2=0$ the system becomes a stack of 
non-interacting two-dimensional planes, and $T_C$ should vanish. 
This limit lies outside the range of applicability of the HTE, and we postpone its
study to future works. Here we mention only that magnetic anisotropy,
which is neglected in our study, should act in the opposite 
direction, i.e., it should enhance the $T_C$ as 
it depresses spin fluctuations. 

Figure \ref{hte}d shows that the ordering temperature decreases very slowly 
as $L$ is increased. Note that the superstructure parameter
values $L= 1, 2, 3, 4, 5, 6$ correspond to the following 
concentrations of magnetic ions: $x=0.33, 0.18, 0.12, 0.09, 0.07, 0.06$. 
To get a closer relation to experiments,
we may consider, e.g., ZnO:Mn,Co,  where the in-plain superexchange values are
$J_1/k_B \sim 50$~K \cite{remark1,Gratens04,DAmbrosio12} and 
$T_s=J_1S(S+1)/k_B \sim 438(188)$~K for $S=5/2(3/2)$. For other
Co-doped semiconductors 66~K $\lesssim J_1/k_B\lesssim 100$~K
\cite{Giebultowicz90,Shapira02,Savoyant14} (and references therein),
i.e., $T_s$ lies within the interval 248~K $\lesssim T_s\lesssim 375$~K. 
The Mn-doped
semiconductors have 
12~K $\lesssim J_1/k_B\lesssim 32$~K \cite{Foner89,Shapira02}, and
105~K $\lesssim T_s\lesssim 280$~K.

Thus, a very 
diluted system may have an appreciable ordering temperature 
($T_C\gtrsim 100$~K)
provided that the
magnetic ions are arranged in a Lieb-Mattis ferrimagnetic superstructure.

In many aspects, the behavior of a ferrimagnet in its ordered state is
similar to that of a ferromagnet with the same value of 
spontaneous magnetization $M_s$. But in a high magnetic field the ferrimagnet
exhibits a transition accompanied by reorientation of its 
sublattices \cite{Tyablikov56,*Tyablikov,Schloemann60,Clark68}. 
At $T=0$ the magnetization per spin has a constant value, 
$\mu _{M,s}=g\mu _BS|n_A-n_B|/(n_A+n_B)$, up to a certain 
critical field, $H_{c,1}$; 
then it grows up linearly to the saturation value, 
$\mu _{M,max}=g\mu _BS$, which is reached 
at a second critical field, $H_{c,2}$. For a two-sublattice ferrimagnet 
having the structure shown in Fig.\ \ref{hon2d}a ($L=1$) 
and $J_1=J_2=J$ we find
$g\mu _BH_{c,1}=JS$, and $H_{c,2}=5H_{c,1}$.
For $J/k_B \sim 20$~K this gives $H_{c,1}\sim 37$~T, $H_{c,2}\sim 185$~T.

The complexes shown in Fig.\ \ref{tetram}b,c may be arranged in many kinds
of networks, to form Lieb-Mattis ferrimagnetic superstructures 
in various 
host semiconductors. 
Figure \ref{hon2d}d shows an example of a 2D 
square superstructure unit cell with $L=2$, which is possible in a cubic host.
It has $n_A=1+2(L-1)$ and $n_B=4(L-1)+2L$.
One can also imagine a 3D cubic network; then Fig.\ \ref{hon2d}d 
corresponds to a face of the 
cubic unit cell having $n_A=1+3(L-1)$, $n_B=3L+12(L-1)$, and the 
concentration of magnetic ions $x=(n_A+n_B)/n_c=(9L-7)/(4L^3)$. 
Formation of such superstructures is possible in perovskite solid 
solutions, like KMn$_x$Mg$_{1-x}$F$_3$ \cite{DAriano82,Breed70}, or
in solutions of multiferroics
PbFe$_{1/2}$Nb$_{1/2}$O$_{3}$ or PbFe$_{1/2}$Ta$_{1/2}$O$_{3}$
with ferroelectric perovskites 
\cite{Sanchez13,Evans13,Laguta13,Sanchez11,Kumar09}.

We conclude that Lieb-Mattis ferrimagnetism is a possible route to obtaining
long-range magnetic order 
in semiconductors containing transition metal ions as substitutional 
impurities, which requires no 
additional charge carriers. A precursor of the ordering transition 
is the enhanced magnetic response
of finite cluster showing isotropic superparamagnetism. Our results for the inverse 
susceptibility show a characteristic "S"-like form of the curves, which could be used to identify the present 
mechanism. Adding the magnetic anisotropy to our theory, we expect also other ingredients 
of superparamagnetism, namely a finite blocking temperature and hysteresis.

These superparamagnetic clusters serve as building blocks to create infinite sublattices of the wurtzite 
structure that obey the Lieb-Mattis rules. As we have already noted, there is an enormous wealth of such
Lieb-Mattis sublattices, our proposals (Fig.\ \ref{hon2d}) may only serve as examples. We expect a finite
transition temperature for all these lattices and we have shown it explicitly for the subclass that we 
considered. Of course, a question arises, whether frustration in a realistic diluted 
semiconductor can influence the above discussed scenario.
First we argue that there are several numerical studies showing that the Lieb-Mattis
theorem, although not rigorously valid, applies to many 
frustrated spin systems, see, e.g., Ref.~\onlinecite{LM_frust}.
Furthermore, we know that there are various frustrated  2D lattices with
antiferromagnetic
nearest-neighbor exchange, such as the triangular or the
Shastry-Sutherland lattices, which show ground-state magnetic LRO
\cite{Richter04,Archi2014}.   
Last but not least, the stability of the ferrimagnetic ground state against
frustration has been demonstrated for several specific ferrimagnetic models,
see, e.g., Refs.~\onlinecite{Ivanov1998,waldtmann2000,Ivanov2002}.  
Consequently, there is ample evidence that the above sketched mechanism 
should be robust against
frustration. 
The final proof that the here proposed mechanism can, indeed, be 
realized in a real material demands further studies, in close collaboration between experiment and theory.

In this communication, we have considered only semiconductors doped by one
kind of magnetic ions, where ferrimagnetism can appear due to the 
topology of interacting bonds. Another option is the co-doping with 
two kinds of ions having different spin values. In both cases
a  ferrimagnetic semiconductor may be a good alternative to 
a ferromagnetic one.

\begin{acknowledgments} 
The projects NASc of Ukraine 07-02-15, and NATO  project SfP 984735
are acknowledged. The exact diagonalization calculations were performed 
using J. Schulenburg's {\it spinpack}.
\end{acknowledgments}

 \bibliographystyle{apsrev}
 \bibliography{wf4} 

\begin{thebibliography}{46}
\expandafter\ifx\csname natexlab\endcsname\relax\def\natexlab#1{#1}\fi
\expandafter\ifx\csname bibnamefont\endcsname\relax
  \def\bibnamefont#1{#1}\fi
\expandafter\ifx\csname bibfnamefont\endcsname\relax
  \def\bibfnamefont#1{#1}\fi
\expandafter\ifx\csname citenamefont\endcsname\relax
  \def\citenamefont#1{#1}\fi
\expandafter\ifx\csname url\endcsname\relax
  \def\url#1{\texttt{#1}}\fi
\expandafter\ifx\csname urlprefix\endcsname\relax\def\urlprefix{URL }\fi
\providecommand{\bibinfo}[2]{#2}
\providecommand{\eprint}[2][]{\url{#2}}

\bibitem[{\citenamefont{\v{Z}uti\'{c} et~al.}(2004)\citenamefont{\v{Z}uti\'{c},
  Fabian, and Das~Sarma}}]{Zutic04}
\bibinfo{author}{\bibfnamefont{I.}~\bibnamefont{\v{Z}uti\'{c}}},
  \bibinfo{author}{\bibfnamefont{J.}~\bibnamefont{Fabian}}, \bibnamefont{and}
  \bibinfo{author}{\bibfnamefont{S.}~\bibnamefont{Das~Sarma}},
  \bibinfo{journal}{Rev. Mod. Phys.} \textbf{\bibinfo{volume}{76}},
  \bibinfo{pages}{323} (\bibinfo{year}{2004}),
  \urlprefix\url{http://link.aps.org/doi/10.1103/RevModPhys.76.323}.

\bibitem[{\citenamefont{Matsukura et~al.}(1998)\citenamefont{Matsukura, Ohno,
  Shen, and Sugawara}}]{Matsukura98}
\bibinfo{author}{\bibfnamefont{F.}~\bibnamefont{Matsukura}},
  \bibinfo{author}{\bibfnamefont{H.}~\bibnamefont{Ohno}},
  \bibinfo{author}{\bibfnamefont{A.}~\bibnamefont{Shen}}, \bibnamefont{and}
  \bibinfo{author}{\bibfnamefont{Y.}~\bibnamefont{Sugawara}},
  \bibinfo{journal}{Phys. Rev. B} \textbf{\bibinfo{volume}{57}},
  \bibinfo{pages}{R2037} (\bibinfo{year}{1998}).

\bibitem[{\citenamefont{Dietl et~al.}(2000)\citenamefont{Dietl, Ohno,
  Matsukura, Cibert, and Ferrand}}]{Dietl00}
\bibinfo{author}{\bibfnamefont{T.}~\bibnamefont{Dietl}},
  \bibinfo{author}{\bibfnamefont{H.}~\bibnamefont{Ohno}},
  \bibinfo{author}{\bibfnamefont{F.}~\bibnamefont{Matsukura}},
  \bibinfo{author}{\bibfnamefont{J.}~\bibnamefont{Cibert}}, \bibnamefont{and}
  \bibinfo{author}{\bibfnamefont{D.}~\bibnamefont{Ferrand}},
  \bibinfo{journal}{Science} \textbf{\bibinfo{volume}{287}},
  \bibinfo{pages}{1019} (\bibinfo{year}{2000}).

\bibitem[{\citenamefont{Zener}(1951)}]{Zener51}
\bibinfo{author}{\bibfnamefont{C.}~\bibnamefont{Zener}},
  \bibinfo{journal}{Phys. Rev.} \textbf{\bibinfo{volume}{82}},
  \bibinfo{pages}{403} (\bibinfo{year}{1951}).

\bibitem[{\citenamefont{Dietl}(2010)}]{Dietl10}
\bibinfo{author}{\bibfnamefont{T.}~\bibnamefont{Dietl}}, \bibinfo{journal}{Nat.
  Mater.} \textbf{\bibinfo{volume}{9}}, \bibinfo{pages}{965}
  (\bibinfo{year}{2010}), \urlprefix\url{http://dx.doi.org/10.1038/nmat2898}.

\bibitem[{\citenamefont{Janisch et~al.}(2005)\citenamefont{Janisch, Gopal, and
  Spaldin}}]{Janisch05}
\bibinfo{author}{\bibfnamefont{R.}~\bibnamefont{Janisch}},
  \bibinfo{author}{\bibfnamefont{P.}~\bibnamefont{Gopal}}, \bibnamefont{and}
  \bibinfo{author}{\bibfnamefont{N.~A.} \bibnamefont{Spaldin}},
  \bibinfo{journal}{Journal of Physics: Condensed Matter}
  \textbf{\bibinfo{volume}{17}}, \bibinfo{pages}{R657} (\bibinfo{year}{2005}),
  \urlprefix\url{http://stacks.iop.org/0953-8984/17/i=27/a=R01}.

\bibitem[{\citenamefont{Ogale}(2010)}]{Ogale10}
\bibinfo{author}{\bibfnamefont{S.~B.} \bibnamefont{Ogale}},
  \bibinfo{journal}{Advanced Materials} \textbf{\bibinfo{volume}{22}},
  \bibinfo{pages}{3125} (\bibinfo{year}{2010}), ISSN \bibinfo{issn}{1521-4095},
  \urlprefix\url{http://dx.doi.org/10.1002/adma.200903891}.

\bibitem[{\citenamefont{Lieb and Mattis}(1962)}]{Lieb62}
\bibinfo{author}{\bibfnamefont{E.}~\bibnamefont{Lieb}} \bibnamefont{and}
  \bibinfo{author}{\bibfnamefont{D.}~\bibnamefont{Mattis}},
  \bibinfo{journal}{Journal of Mathematical Physics}
  \textbf{\bibinfo{volume}{3}}, \bibinfo{pages}{749} (\bibinfo{year}{1962}),
  \urlprefix\url{http://link.aip.org/link/?JMP/3/749/1}.

\bibitem[{\citenamefont{Bedanta and Kleemann}(2009)}]{Bedanta09}
\bibinfo{author}{\bibfnamefont{S.}~\bibnamefont{Bedanta}} \bibnamefont{and}
  \bibinfo{author}{\bibfnamefont{W.}~\bibnamefont{Kleemann}},
  \bibinfo{journal}{Journal of Physics D: Applied Physics}
  \textbf{\bibinfo{volume}{42}}, \bibinfo{pages}{013001}
  (\bibinfo{year}{2009}),
  \urlprefix\url{http://stacks.iop.org/0022-3727/42/i=1/a=013001}.

\bibitem[{\citenamefont{Kuzian et~al.}(2014)\citenamefont{Kuzian, Laguta, and
  Richter}}]{Kuzian14b}
\bibinfo{author}{\bibfnamefont{R.~O.} \bibnamefont{Kuzian}},
  \bibinfo{author}{\bibfnamefont{V.~V.} \bibnamefont{Laguta}},
  \bibnamefont{and} \bibinfo{author}{\bibfnamefont{J.}~\bibnamefont{Richter}},
  \bibinfo{journal}{Phys. Rev. B} \textbf{\bibinfo{volume}{90}},
  \bibinfo{pages}{134415} (\bibinfo{year}{2014}),
  \urlprefix\url{http://link.aps.org/doi/10.1103/PhysRevB.90.134415}.

\bibitem[{rem({\natexlab{a}})}]{remark1}
\bibinfo{note}{In the literature one meets also the notation
  $-2J_{\mathbf{r},L}$ for the same exchange parameter.}

\bibitem[{\citenamefont{Shapira and Bindilatti}(2002)}]{Shapira02}
\bibinfo{author}{\bibfnamefont{Y.}~\bibnamefont{Shapira}} \bibnamefont{and}
  \bibinfo{author}{\bibfnamefont{V.}~\bibnamefont{Bindilatti}},
  \bibinfo{journal}{Journal of Applied Physics} \textbf{\bibinfo{volume}{92}},
  \bibinfo{pages}{4155} (\bibinfo{year}{2002}),
  \urlprefix\url{http://scitation.aip.org/content/aip/journal/jap/92/8/10.1063/1.1507808}.

\bibitem[{\citenamefont{Gratens et~al.}(2004)\citenamefont{Gratens, Bindilatti,
  Oliveira, Shapira, Foner, Golacki, and Haas}}]{Gratens04}
\bibinfo{author}{\bibfnamefont{X.}~\bibnamefont{Gratens}},
  \bibinfo{author}{\bibfnamefont{V.}~\bibnamefont{Bindilatti}},
  \bibinfo{author}{\bibfnamefont{N.~F.} \bibnamefont{Oliveira}},
  \bibinfo{author}{\bibfnamefont{Y.}~\bibnamefont{Shapira}},
  \bibinfo{author}{\bibfnamefont{S.}~\bibnamefont{Foner}},
  \bibinfo{author}{\bibfnamefont{Z.}~\bibnamefont{Golacki}}, \bibnamefont{and}
  \bibinfo{author}{\bibfnamefont{T.~E.} \bibnamefont{Haas}},
  \bibinfo{journal}{Phys. Rev. B} \textbf{\bibinfo{volume}{69}},
  \bibinfo{pages}{125209} (\bibinfo{year}{2004}),
  \urlprefix\url{http://link.aps.org/doi/10.1103/PhysRevB.69.125209}.

\bibitem[{\citenamefont{D'Ambrosio et~al.}(2012)\citenamefont{D'Ambrosio,
  Pashchenko, Mignot, Ignatchik, Kuzian, Savoyant, Golacki, Grasza, and
  Stepanov}}]{DAmbrosio12}
\bibinfo{author}{\bibfnamefont{S.}~\bibnamefont{D'Ambrosio}},
  \bibinfo{author}{\bibfnamefont{V.}~\bibnamefont{Pashchenko}},
  \bibinfo{author}{\bibfnamefont{J.-M.} \bibnamefont{Mignot}},
  \bibinfo{author}{\bibfnamefont{O.}~\bibnamefont{Ignatchik}},
  \bibinfo{author}{\bibfnamefont{R.~O.} \bibnamefont{Kuzian}},
  \bibinfo{author}{\bibfnamefont{A.}~\bibnamefont{Savoyant}},
  \bibinfo{author}{\bibfnamefont{Z.}~\bibnamefont{Golacki}},
  \bibinfo{author}{\bibfnamefont{K.}~\bibnamefont{Grasza}}, \bibnamefont{and}
  \bibinfo{author}{\bibfnamefont{A.}~\bibnamefont{Stepanov}},
  \bibinfo{journal}{Phys. Rev. B} \textbf{\bibinfo{volume}{86}},
  \bibinfo{pages}{035202} (\bibinfo{year}{2012}),
  \urlprefix\url{http://link.aps.org/doi/10.1103/PhysRevB.86.035202}.

\bibitem[{\citenamefont{Chanier et~al.}(2006)\citenamefont{Chanier, Sargolzaei,
  Opahle, Hayn, and Koepernik}}]{Chanier06}
\bibinfo{author}{\bibfnamefont{T.}~\bibnamefont{Chanier}},
  \bibinfo{author}{\bibfnamefont{M.}~\bibnamefont{Sargolzaei}},
  \bibinfo{author}{\bibfnamefont{I.}~\bibnamefont{Opahle}},
  \bibinfo{author}{\bibfnamefont{R.}~\bibnamefont{Hayn}}, \bibnamefont{and}
  \bibinfo{author}{\bibfnamefont{K.}~\bibnamefont{Koepernik}},
  \bibinfo{journal}{Phys. Rev. B} \textbf{\bibinfo{volume}{73}},
  \bibinfo{pages}{134418} (\bibinfo{year}{2006}).

\bibitem[{\citenamefont{Kuzian et~al.}(2011)\citenamefont{Kuzian, Dar\'e,
  Savoyant, D'Ambrosio, and Stepanov}}]{Kuzian11}
\bibinfo{author}{\bibfnamefont{R.~O.} \bibnamefont{Kuzian}},
  \bibinfo{author}{\bibfnamefont{A.~M.} \bibnamefont{Dar\'e}},
  \bibinfo{author}{\bibfnamefont{A.}~\bibnamefont{Savoyant}},
  \bibinfo{author}{\bibfnamefont{S.}~\bibnamefont{D'Ambrosio}},
  \bibnamefont{and} \bibinfo{author}{\bibfnamefont{A.}~\bibnamefont{Stepanov}},
  \bibinfo{journal}{Phys. Rev. B} \textbf{\bibinfo{volume}{84}},
  \bibinfo{pages}{165207} (\bibinfo{year}{2011}),
  \urlprefix\url{http://link.aps.org/doi/10.1103/PhysRevB.84.165207}.

\bibitem[{\citenamefont{Savoyant et~al.}(2014)\citenamefont{Savoyant,
  D'Ambrosio, Kuzian, Dar\'e, and Stepanov}}]{Savoyant14}
\bibinfo{author}{\bibfnamefont{A.}~\bibnamefont{Savoyant}},
  \bibinfo{author}{\bibfnamefont{S.}~\bibnamefont{D'Ambrosio}},
  \bibinfo{author}{\bibfnamefont{R.~O.} \bibnamefont{Kuzian}},
  \bibinfo{author}{\bibfnamefont{A.~M.} \bibnamefont{Dar\'e}},
  \bibnamefont{and} \bibinfo{author}{\bibfnamefont{A.}~\bibnamefont{Stepanov}},
  \bibinfo{journal}{Phys. Rev. B} \textbf{\bibinfo{volume}{90}},
  \bibinfo{pages}{075205} (\bibinfo{year}{2014}),
  \urlprefix\url{http://link.aps.org/doi/10.1103/PhysRevB.90.075205}.

\bibitem[{\citenamefont{Liu et~al.}(1996)\citenamefont{Liu, Shapira, Haar,
  Bindilatti, and McNiff}}]{Liu96}
\bibinfo{author}{\bibfnamefont{M.~T.} \bibnamefont{Liu}},
  \bibinfo{author}{\bibfnamefont{Y.}~\bibnamefont{Shapira}},
  \bibinfo{author}{\bibfnamefont{E.} \bibnamefont{ter Haar}},
  \bibinfo{author}{\bibfnamefont{V.}~\bibnamefont{Bindilatti}},
  \bibnamefont{and} \bibinfo{author}{\bibfnamefont{E.~J.}
  \bibnamefont{McNiff}}, \bibinfo{journal}{Phys. Rev. B}
  \textbf{\bibinfo{volume}{54}}, \bibinfo{pages}{6457} (\bibinfo{year}{1996}),
  \urlprefix\url{http://link.aps.org/doi/10.1103/PhysRevB.54.6457}.

\bibitem[{rem({\natexlab{b}})}]{remark2}
\bibinfo{note}{The term ``ferri-magnetism'' for finite system means that, on
  one hand, quantum mechanical average over the ground state of operators of
  neighboring spins $\langle
  \hat{S}_{\mathbf{R}}\hat{S}_{\mathbf{R+\rho}}\rangle$ (vector $\rho $ connect
  neighboring spin positions)is negative (in average, the neighboring spins are
  aligned in opposite directions), whereas, on the other hand, the total ground
  state spin of the system $S_g$ is non-zero.}

\bibitem[{spi()}]{spinpack1}
\bibinfo{note}{{\it spinpack} is available at
  http://www-e.uni-magdeburg.de/jschulen/spin/}.

\bibitem[{\citenamefont{Richter and Schulenburg}(2010)}]{spinpack2}
\bibinfo{author}{\bibfnamefont{J.}~\bibnamefont{Richter}} \bibnamefont{and}
  \bibinfo{author}{\bibfnamefont{J.}~\bibnamefont{Schulenburg}},
  \bibinfo{journal}{Eur. Phys. J. B} \textbf{\bibinfo{volume}{73}},
  \bibinfo{pages}{117} (\bibinfo{year}{2010}).

\bibitem[{\citenamefont{Laguta et~al.}(2014)\citenamefont{Laguta, Stephanovich,
  Savinov, Marysko, Kuzian, Kondakova, Olekhnovich, Pushkarev, Radyush, Raevski
  et~al.}}]{Laguta14}
\bibinfo{author}{\bibfnamefont{V.~V.} \bibnamefont{Laguta}},
  \bibinfo{author}{\bibfnamefont{V.~A.} \bibnamefont{Stephanovich}},
  \bibinfo{author}{\bibfnamefont{M.}~\bibnamefont{Savinov}},
  \bibinfo{author}{\bibfnamefont{M.}~\bibnamefont{Marysko}},
  \bibinfo{author}{\bibfnamefont{R.~O.} \bibnamefont{Kuzian}},
  \bibinfo{author}{\bibfnamefont{I.~V.} \bibnamefont{Kondakova}},
  \bibinfo{author}{\bibfnamefont{N.~M.} \bibnamefont{Olekhnovich}},
  \bibinfo{author}{\bibfnamefont{A.~V.} \bibnamefont{Pushkarev}},
  \bibinfo{author}{\bibfnamefont{Y.~V.} \bibnamefont{Radyush}},
  \bibinfo{author}{\bibfnamefont{I.~P.} \bibnamefont{Raevski}},
  \bibnamefont{et~al.}, \bibinfo{journal}{New Journal of Physics}
  \textbf{\bibinfo{volume}{16}}, \bibinfo{pages}{113041}
  (\bibinfo{year}{2014}),
  \urlprefix\url{http://stacks.iop.org/1367-2630/16/i=11/a=113041}.

\bibitem[{\citenamefont{Schmidt et~al.}(2011)\citenamefont{Schmidt, Lohmann,
  and Richter}}]{Schmidt11}
\bibinfo{author}{\bibfnamefont{H.-J.} \bibnamefont{Schmidt}},
  \bibinfo{author}{\bibfnamefont{A.}~\bibnamefont{Lohmann}}, \bibnamefont{and}
  \bibinfo{author}{\bibfnamefont{J.}~\bibnamefont{Richter}},
  \bibinfo{journal}{Phys. Rev. B} \textbf{\bibinfo{volume}{84}},
  \bibinfo{pages}{104443} (\bibinfo{year}{2011}),
  \urlprefix\url{http://link.aps.org/doi/10.1103/PhysRevB.84.104443}.

\bibitem[{hte({\natexlab{a}})}]{hte8}
\bibinfo{note}{For eight-order HTE, we have used the 2011-09-23 version of HTE
  package available at http://www.uni-magdeburg.de/jschulen/HTE/},
  \urlprefix\url{http://www.uni-magdeburg.de/jschulen/HTE/}.

\bibitem[{\citenamefont{Lohmann et~al.}(2014)\citenamefont{Lohmann, Schmidt,
  and Richter}}]{Lohmann14}
\bibinfo{author}{\bibfnamefont{A.}~\bibnamefont{Lohmann}},
  \bibinfo{author}{\bibfnamefont{H.-J.} \bibnamefont{Schmidt}},
  \bibnamefont{and} \bibinfo{author}{\bibfnamefont{J.}~\bibnamefont{Richter}},
  \bibinfo{journal}{Phys. Rev. B} \textbf{\bibinfo{volume}{89}},
  \bibinfo{pages}{014415} (\bibinfo{year}{2014}),
  \urlprefix\url{http://link.aps.org/doi/10.1103/PhysRevB.89.014415}.

\bibitem[{Loh()}]{Lohmann}
\bibinfo{note}{We thank A. Lohmann for providing the code of the 11th order
  HTE.}

\bibitem[{hte({\natexlab{b}})}]{hte10}
\bibinfo{note}{For tenth-order HTE, we have used HTE10 package available at
  http://www.uni-magdeburg.de/jschulen/HTE10/},
  \urlprefix\url{http://www.uni-magdeburg.de/jschulen/HTE10/}.

\bibitem[{\citenamefont{Giebultowicz et~al.}(1990)\citenamefont{Giebultowicz,
  Rhyne, Furdyna, and Klosowski}}]{Giebultowicz90}
\bibinfo{author}{\bibfnamefont{T.~M.} \bibnamefont{Giebultowicz}},
  \bibinfo{author}{\bibfnamefont{J.~J.} \bibnamefont{Rhyne}},
  \bibinfo{author}{\bibfnamefont{J.~K.} \bibnamefont{Furdyna}},
  \bibnamefont{and}
  \bibinfo{author}{\bibfnamefont{P.}~\bibnamefont{Klosowski}},
  \bibinfo{journal}{Journal of Applied Physics} \textbf{\bibinfo{volume}{67}},
  \bibinfo{pages}{5096} (\bibinfo{year}{1990}),
  \urlprefix\url{http://scitation.aip.org/content/aip/journal/jap/67/9/10.1063/1.344683}.

\bibitem[{\citenamefont{Foner et~al.}(1989)\citenamefont{Foner, Shapira,
  Heiman, Becla, Kershaw, Dwight, and Wold}}]{Foner89}
\bibinfo{author}{\bibfnamefont{S.}~\bibnamefont{Foner}},
  \bibinfo{author}{\bibfnamefont{Y.}~\bibnamefont{Shapira}},
  \bibinfo{author}{\bibfnamefont{D.}~\bibnamefont{Heiman}},
  \bibinfo{author}{\bibfnamefont{P.}~\bibnamefont{Becla}},
  \bibinfo{author}{\bibfnamefont{R.}~\bibnamefont{Kershaw}},
  \bibinfo{author}{\bibfnamefont{K.}~\bibnamefont{Dwight}}, \bibnamefont{and}
  \bibinfo{author}{\bibfnamefont{A.}~\bibnamefont{Wold}},
  \bibinfo{journal}{Phys. Rev. B} \textbf{\bibinfo{volume}{39}},
  \bibinfo{pages}{11793} (\bibinfo{year}{1989}),
  \urlprefix\url{http://link.aps.org/doi/10.1103/PhysRevB.39.11793}.

\bibitem[{\citenamefont{Tyablikov}(1956)}]{Tyablikov56}
\bibinfo{author}{\bibfnamefont{S.~V.} \bibnamefont{Tyablikov}},
  \bibinfo{journal}{Fiz. Metallov. i Metallovedenie}
  \textbf{\bibinfo{volume}{3}}, \bibinfo{pages}{3} (\bibinfo{year}{1956}).

\bibitem[{\citenamefont{Tyablikov}(1967)}]{Tyablikov}
\bibinfo{author}{\bibfnamefont{S.~V.} \bibnamefont{Tyablikov}},
  \emph{\bibinfo{title}{Methods in the Quantum Theory of Magnetism}}
  (\bibinfo{publisher}{Plenum}, \bibinfo{address}{New York},
  \bibinfo{year}{1967}).

\bibitem[{\citenamefont{Schl\"omann}(1960)}]{Schloemann60}
\bibinfo{author}{\bibfnamefont{E.}~\bibnamefont{Schl\"omann}}, in
  \emph{\bibinfo{booktitle}{Solid State Physics in Electronics and
  Telecommunications}}, edited by
  \bibinfo{editor}{\bibfnamefont{M.}~\bibnamefont{D\'esirant}}
  \bibnamefont{and} \bibinfo{editor}{\bibfnamefont{J.}~\bibnamefont{Michiels}}
  (\bibinfo{publisher}{Academic Press, London}, \bibinfo{year}{1960}).

\bibitem[{\citenamefont{Clark and Callen}(1968)}]{Clark68}
\bibinfo{author}{\bibfnamefont{A.~E.} \bibnamefont{Clark}} \bibnamefont{and}
  \bibinfo{author}{\bibfnamefont{E.}~\bibnamefont{Callen}},
  \bibinfo{journal}{Journal of Applied Physics} \textbf{\bibinfo{volume}{39}},
  \bibinfo{pages}{5972} (\bibinfo{year}{1968}),
  \urlprefix\url{http://scitation.aip.org/content/aip/journal/jap/39/13/10.1063/1.1656100}.

\bibitem[{\citenamefont{D'Ariano and Borsa}(1982)}]{DAriano82}
\bibinfo{author}{\bibfnamefont{G.}~\bibnamefont{D'Ariano}} \bibnamefont{and}
  \bibinfo{author}{\bibfnamefont{F.}~\bibnamefont{Borsa}},
  \bibinfo{journal}{Phys. Rev. B} \textbf{\bibinfo{volume}{26}},
  \bibinfo{pages}{6215} (\bibinfo{year}{1982}),
  \urlprefix\url{http://link.aps.org/doi/10.1103/PhysRevB.26.6215}.

\bibitem[{\citenamefont{Breed et~al.}(1970)\citenamefont{Breed, Gilijamse,
  Sterkenburg, and Miedema}}]{Breed70}
\bibinfo{author}{\bibfnamefont{D.~J.} \bibnamefont{Breed}},
  \bibinfo{author}{\bibfnamefont{K.}~\bibnamefont{Gilijamse}},
  \bibinfo{author}{\bibfnamefont{J.~W.~E.} \bibnamefont{Sterkenburg}},
  \bibnamefont{and} \bibinfo{author}{\bibfnamefont{A.~R.}
  \bibnamefont{Miedema}}, \bibinfo{journal}{J. Appl. Phys.}
  \textbf{\bibinfo{volume}{41}}, \bibinfo{pages}{1267} (\bibinfo{year}{1970}),
  \urlprefix\url{http://dx.doi.org.sci-hub.org/10.1063/1.1658906}.

\bibitem[{\citenamefont{Sanchez et~al.}(2013)\citenamefont{Sanchez, Ortega,
  Kumar, Sreenivasulu, Katiyar, Scott, Evans, Arredondo-Arechavala, Schilling,
  and Gregg}}]{Sanchez13}
\bibinfo{author}{\bibfnamefont{D.~A.} \bibnamefont{Sanchez}},
  \bibinfo{author}{\bibfnamefont{N.}~\bibnamefont{Ortega}},
  \bibinfo{author}{\bibfnamefont{A.}~\bibnamefont{Kumar}},
  \bibinfo{author}{\bibfnamefont{G.}~\bibnamefont{Sreenivasulu}},
  \bibinfo{author}{\bibfnamefont{R.~S.} \bibnamefont{Katiyar}},
  \bibinfo{author}{\bibfnamefont{J.~F.} \bibnamefont{Scott}},
  \bibinfo{author}{\bibfnamefont{D.~M.} \bibnamefont{Evans}},
  \bibinfo{author}{\bibfnamefont{M.}~\bibnamefont{Arredondo-Arechavala}},
  \bibinfo{author}{\bibfnamefont{A.}~\bibnamefont{Schilling}},
  \bibnamefont{and} \bibinfo{author}{\bibfnamefont{J.~M.} \bibnamefont{Gregg}},
  \bibinfo{journal}{Journal of Applied Physics} \textbf{\bibinfo{volume}{113}},
  \bibinfo{pages}{074105} (\bibinfo{year}{2013}),
  \urlprefix\url{http://link.aip.org/link/?JAP/113/074105/1}.

\bibitem[{\citenamefont{Evans et~al.}(2013)\citenamefont{Evans, Schilling,
  Kumar, Sanchez, Ortega, Arredondo, Katiyar, Gregg, and Scott}}]{Evans13}
\bibinfo{author}{\bibfnamefont{D.}~\bibnamefont{Evans}},
  \bibinfo{author}{\bibfnamefont{A.}~\bibnamefont{Schilling}},
  \bibinfo{author}{\bibfnamefont{A.}~\bibnamefont{Kumar}},
  \bibinfo{author}{\bibfnamefont{D.}~\bibnamefont{Sanchez}},
  \bibinfo{author}{\bibfnamefont{N.}~\bibnamefont{Ortega}},
  \bibinfo{author}{\bibfnamefont{M.}~\bibnamefont{Arredondo}},
  \bibinfo{author}{\bibfnamefont{R.}~\bibnamefont{Katiyar}},
  \bibinfo{author}{\bibfnamefont{J.}~\bibnamefont{Gregg}}, \bibnamefont{and}
  \bibinfo{author}{\bibfnamefont{J.}~\bibnamefont{Scott}},
  \bibinfo{journal}{Nat. Commun.} \textbf{\bibinfo{volume}{4}},
  \bibinfo{pages}{1534} (\bibinfo{year}{2013}),
  \urlprefix\url{http://dx.doi.org/10.1038/ncomms2548}.

\bibitem[{\citenamefont{Laguta et~al.}(2013)\citenamefont{Laguta, Glinchuk,
  Mary\v{s}ko, Kuzian, Prosandeev, Raevskaya, Smotrakov, Eremkin, and
  Raevski}}]{Laguta13}
\bibinfo{author}{\bibfnamefont{V.~V.} \bibnamefont{Laguta}},
  \bibinfo{author}{\bibfnamefont{M.~D.} \bibnamefont{Glinchuk}},
  \bibinfo{author}{\bibfnamefont{M.}~\bibnamefont{Mary\v{s}ko}},
  \bibinfo{author}{\bibfnamefont{R.~O.} \bibnamefont{Kuzian}},
  \bibinfo{author}{\bibfnamefont{S.~A.} \bibnamefont{Prosandeev}},
  \bibinfo{author}{\bibfnamefont{S.~I.} \bibnamefont{Raevskaya}},
  \bibinfo{author}{\bibfnamefont{V.~G.} \bibnamefont{Smotrakov}},
  \bibinfo{author}{\bibfnamefont{V.~V.} \bibnamefont{Eremkin}},
  \bibnamefont{and} \bibinfo{author}{\bibfnamefont{I.~P.}
  \bibnamefont{Raevski}}, \bibinfo{journal}{Phys. Rev. B}
  \textbf{\bibinfo{volume}{87}}, \bibinfo{pages}{064403}
  (\bibinfo{year}{2013}),
  \urlprefix\url{http://link.aps.org/doi/10.1103/PhysRevB.87.064403}.

\bibitem[{\citenamefont{Sanchez et~al.}(2011)\citenamefont{Sanchez, Ortega,
  Kumar, Roque-Malherbe, Polanco, Scott, and Katiyar}}]{Sanchez11}
\bibinfo{author}{\bibfnamefont{D.~A.} \bibnamefont{Sanchez}},
  \bibinfo{author}{\bibfnamefont{N.}~\bibnamefont{Ortega}},
  \bibinfo{author}{\bibfnamefont{A.}~\bibnamefont{Kumar}},
  \bibinfo{author}{\bibfnamefont{R.}~\bibnamefont{Roque-Malherbe}},
  \bibinfo{author}{\bibfnamefont{R.}~\bibnamefont{Polanco}},
  \bibinfo{author}{\bibfnamefont{J.~F.} \bibnamefont{Scott}}, \bibnamefont{and}
  \bibinfo{author}{\bibfnamefont{R.~S.} \bibnamefont{Katiyar}},
  \bibinfo{journal}{AIP Advances} \textbf{\bibinfo{volume}{1}},
  \bibinfo{pages}{042169} (\bibinfo{year}{2011}),
  \urlprefix\url{http://link.aip.org/link/?ADV/1/042169/1}.

\bibitem[{\citenamefont{Kumar et~al.}(2009)\citenamefont{Kumar, Sharma,
  Katiyar, Pirc, Blinc, and Scott}}]{Kumar09}
\bibinfo{author}{\bibfnamefont{A.}~\bibnamefont{Kumar}},
  \bibinfo{author}{\bibfnamefont{G.~L.} \bibnamefont{Sharma}},
  \bibinfo{author}{\bibfnamefont{R.~S.} \bibnamefont{Katiyar}},
  \bibinfo{author}{\bibfnamefont{R.}~\bibnamefont{Pirc}},
  \bibinfo{author}{\bibfnamefont{R.}~\bibnamefont{Blinc}}, \bibnamefont{and}
  \bibinfo{author}{\bibfnamefont{J.~F.} \bibnamefont{Scott}},
  \bibinfo{journal}{Journal of Physics: Condensed Matter}
  \textbf{\bibinfo{volume}{21}}, \bibinfo{pages}{382204}
  (\bibinfo{year}{2009}),
  \urlprefix\url{http://stacks.iop.org/0953-8984/21/i=38/a=382204}.

\bibitem[{\citenamefont{Richter et~al.}(1995)\citenamefont{Richter, Ivanov,
  Retzlaff, and Voigt}}]{LM_frust}
\bibinfo{author}{\bibfnamefont{J.}~\bibnamefont{Richter}},
  \bibinfo{author}{\bibfnamefont{N.}~\bibnamefont{Ivanov}},
  \bibinfo{author}{\bibfnamefont{K.}~\bibnamefont{Retzlaff}}, \bibnamefont{and}
  \bibinfo{author}{\bibfnamefont{A.}~\bibnamefont{Voigt}}, \bibinfo{journal}{J.
  Magn. Magn. Mat.} \textbf{\bibinfo{volume}{140-144}}, \bibinfo{pages}{1611}
  (\bibinfo{year}{1995}).

\bibitem[{\citenamefont{Richter et~al.}(2004)\citenamefont{Richter,
  Schulenburg, and Honecker}}]{Richter04}
\bibinfo{author}{\bibfnamefont{J.}~\bibnamefont{Richter}},
  \bibinfo{author}{\bibfnamefont{J.}~\bibnamefont{Schulenburg}},
  \bibnamefont{and} \bibinfo{author}{\bibfnamefont{A.}~\bibnamefont{Honecker}},
  \bibinfo{journal}{Lect. Notes Phys.} \textbf{\bibinfo{volume}{645}},
  \bibinfo{pages}{85} (\bibinfo{year}{2004}).

\bibitem[{\citenamefont{Farnell et~al.}(2014)\citenamefont{Farnell, G\"otze,
  Richter, Bishop, and Li}}]{Archi2014}
\bibinfo{author}{\bibfnamefont{D.J.J.}~\bibnamefont{Farnell}},
  \bibinfo{author}{\bibfnamefont{O.}~\bibnamefont{G\"otze}},
  \bibinfo{author}{\bibfnamefont{J.}~\bibnamefont{Richter}},
  \bibinfo{author}{\bibfnamefont{R.F.}~\bibnamefont{Bishop}}, \bibnamefont{and}
  \bibinfo{author}{\bibfnamefont{P.H.Y.}~\bibnamefont{Li}}, \bibinfo{journal}{Phys.
  Rev. B} \textbf{\bibinfo{volume}{89}}, \bibinfo{pages}{184407}
  (\bibinfo{year}{2014}).

\bibitem[{\citenamefont{Ivanov et~al.}(1998)\citenamefont{Ivanov, Richter, and
  Schollw\"ock}}]{Ivanov1998}
\bibinfo{author}{\bibfnamefont{N.B.}~\bibnamefont{Ivanov}},
  \bibinfo{author}{\bibfnamefont{J.}~\bibnamefont{Richter}}, \bibnamefont{and}
  \bibinfo{author}{\bibfnamefont{U.}~\bibnamefont{Schollw\"ock}},
  \bibinfo{journal}{Phys. Rev. B} \textbf{\bibinfo{volume}{58}},
  \bibinfo{pages}{14456} (\bibinfo{year}{1998}).

\bibitem[{\citenamefont{Waldtmann et~al.}(2000)\citenamefont{Waldtmann,
  Kreutzmann, Schollw\"ock, Maisinger, and Everts}}]{waldtmann2000}
\bibinfo{author}{\bibfnamefont{C.}~\bibnamefont{Waldtmann}},
  \bibinfo{author}{\bibfnamefont{H.}~\bibnamefont{Kreutzmann}},
  \bibinfo{author}{\bibfnamefont{U.}~\bibnamefont{Schollw\"ock}},
  \bibinfo{author}{\bibfnamefont{K.}~\bibnamefont{Maisinger}},
  \bibnamefont{and} \bibinfo{author}{\bibfnamefont{H.-U.}
  \bibnamefont{Everts}}, \bibinfo{journal}{Phys. Rev. B}
  \textbf{\bibinfo{volume}{62}}, \bibinfo{pages}{9472} (\bibinfo{year}{2000}).

\bibitem[{\citenamefont{Ivanov et~al.}(2002)\citenamefont{Ivanov, Richter, and
  Farnell}}]{Ivanov2002}
\bibinfo{author}{\bibfnamefont{N.B.}~\bibnamefont{Ivanov}},
  \bibinfo{author}{\bibfnamefont{J.}~\bibnamefont{Richter}}, \bibnamefont{and}
  \bibinfo{author}{\bibfnamefont{D.J.J.}~\bibnamefont{Farnell}},
  \bibinfo{journal}{Phys. Rev. B} \textbf{\bibinfo{volume}{66}},
  \bibinfo{pages}{014421} (\bibinfo{year}{2002}).

\end{thebibliography}
\end{document}